\documentclass{PoS}

\title{Precision of analytical approximations in calculations of Atmospheric Leptons}

\ShortTitle{Atmospheric leptons}

\author{\speaker{Thomas K. Gaisser}$^\dagger$, Dennis Soldin, Andrew Crossman\\
        Bartol Research Institute, Dept. of Physics and Astronomy\\
        University of Delaware, Newark, DE 19716 USA\\
        $^\dagger$ E-mail: \email{gaisser@udel.edu}}

\author{Anatoli Fedynitch\\
		Dept.\ of Physics, University of Alberta, Edmonton, Alberta, Canada T6G 2E1\\
        \quad}
        
\abstract{We use the Matrix Cascade Equation code ({\sc MCEq}) to evaluate the range of 
applicability of simple analytic approximations parameterized by spectrum-weighted 
moments and power-law spectral indices that vary slowly with energy.  We compare 
spectra of leptons as a function of zenith angle and energy between MCEq and the 
analytic approximation.  
We also compare fluxes obtained with different models of hadronic interactions.  The goal is to quantify the 
effects of the approximations inherent in the simpler formulas in order to determine 
their limitations and the conditions under which they may be used.  Specifically we look for 
 the range of phase space for which the errors in the approximate formulas 
are smaller than the differences among several different hadronic interaction 
models.  Potential applications include the muon charge ratio, the fraction of prompt 
leptons from decay of charm and seasonal variations of muons and neutrinos.}

\FullConference{36th International Cosmic Ray Conference -ICRC2019-\\
		July 24th - August 1st, 2019\\
		Madison, WI, U.S.A.}
		
\begin{document}

\section{Introduction and formalism}
The production of secondary particles in the atmosphere is governed by the cascade
equation
\begin{equation}
\frac{{\rm d}N_i(E_i,x,\theta)}{{\rm d}X} = -\frac{N_i(E_i,X,\theta)}{\lambda_i(E_i)}-\frac{N_i(E_i,X,\theta)}{d_i(E_i)} + \sum_{j=i}^J\int_{E_i}^\infty\,\left\{\frac {{\rm d}\sigma_{ji}(E_j,E_i)}{\sigma_j^{inel}(E_j)\,\lambda_j(E_j)}+\frac{R_{ji}}{d_j}\right\}N_j(E_j,X,\theta)\,{\rm d}E_j,
\label{eq:MasterEqn}
\end{equation}
where $N_i$ is the flux of particle type $i$, $X$ is the slant depth (g/cm$^2$) in the atmosphere and
the right side of the equation contains the loss terms for interaction and decay and the 
source term for particles of higher energy to produce the secondary particle of interest through interaction or decay.
The decay length has to be converted to g/cm$^2$ and is related to the particle lifetime by
$d_i=\rho(X)\gamma c\tau_i$ by its Lorentz factor $\gamma$ and the density $\rho$ at slant depth $X$.

Solutions of Eq.~(\ref{eq:MasterEqn}) with the boundary condition $N_i(E_0,0,\theta) = \phi_N(E_0)\delta_{iN}$,
where $\phi_N$ is the isotropic flux of primary nucleons, lead 
to inclusive fluxes of secondary particles in the atmosphere.  Solutions of the same equation 
subject to $N_i(E_0,0,\theta) = \delta(E_0 - AE_N)\delta_{iA}$ are air showers.  The focus of
this paper is inclusive fluxes, by which is meant the flux of a particular particle type that would
be measured over a long time period by a detector with acceptance (area-solid angle) so small that
it measures only one particle at a time\footnote{In a real detector, when more than one particle
of a given type is recorded in the same time window, each particle has to be added to the appropriate
bin of energy and particle type.}.

If the production cross sections depend only on the ratio $x=E_i/E_j$ and if the primary spectrum 
of nucleons follows a power law, then explicit solutions of Eq.~(\ref{eq:MasterEqn}) can be obtained 
for the production spectra of atmospheric leptons 
separately in the low- and high-energy limits.  High and low are defined for each channel
in terms of the critical energy at which meson decay and re-interaction are equal.  The solutions are
\begin{equation}
\frac{{\rm d}N_\ell}{{\rm d}E_\ell} =\left . \int_0^{X_0/\cos\theta} P_{M\ell}(E_\ell,X)\,{\rm d}X \right |_{E_\ell\ll\epsilon_M}\rightarrow\frac{N_0(E_\ell)}{1-Z_{NN}}A_{M\ell}
\label{eq:lowE}
\end{equation} 
for low energy and
\begin{equation}
\frac{{\rm d}N_\ell}{{\rm d}E_\ell} =\left . \int_0^{X_0/\cos\theta} P_{M\ell}(E_\ell,X)\,{\rm d}X \right |_{E_\ell\gg\epsilon_M}\rightarrow\frac{N_0(E_\ell)}{1-Z_{NN}}\left(\frac{\epsilon_M}{\cos\theta\,E_\ell}\right)\frac{A_{M\ell}}{B_{M\ell}}
\label{eq:highE}
\end{equation}
for high energy.  The critical energy for meson $M$ is 
\begin{equation}
\epsilon_M=\frac{RT(X)}{M_{\rm mol}g}\frac{Mc^2}{c\tau_M},
\label{eq:critical}
\end{equation}
where T is in $^\circ$K. 
Explicit forms for the lepton production spectra are given in Ref.~\cite{Gaisser:2016uoy}
and in another paper at this conference~\cite{Tilav:2019icrc}.  At $T=220^\circ$, $\epsilon_\pi=115$~GeV and $\epsilon_K=857$~GeV.
The forms for the integrated spectra apply deep in the atmosphere where there is no further
production of leptons of energy $E_\ell$.  The quantities $A$ and $B$ are
\begin{equation}
A_{M\ell} = R_{M\ell}Z_{NM}Z_{M\ell}(\gamma)
\label{eq:Afactor}
\end{equation}
and
\begin{equation}
B_{M\ell} = \frac{Z_{M\ell}(\gamma)}{Z_{M\ell}(\gamma+1)}\frac{\Lambda_M - \Lambda_N}{\Lambda_M\ln(\Lambda_M/\Lambda_N)}.
\label{eq:Bfactor}
\end{equation}

In these equations, $R_{NM}$ is the branching ratio for meson $M$ ($\pi$ or K)  to 
decay to lepton $\ell$ (muon or muon-neutrino), $Z_{NM}$ is the spectrum-weighted moment
for production of meson $M$ in the inelastic collision of a nucleon ($N$) and
$Z_{M\ell}$ is the spectrum weighted moment of the meson decay.  Upper case $\Lambda$'s are
atmospheric attenuation factors that account for the elasticity of a hadron that emerges
from a collision with reduced energy.  In this approximation, the lepton flux is
proportional to the primary spectrum of nucleons $N_0(E_\ell)$ \textit{evaluated at
the energy of the lepton}.  The spectrum weighted moments reflect the relation of
the interaction energy to that of the lepton.

\begin{figure}[b]
\centering
\includegraphics[width=0.5\textwidth]{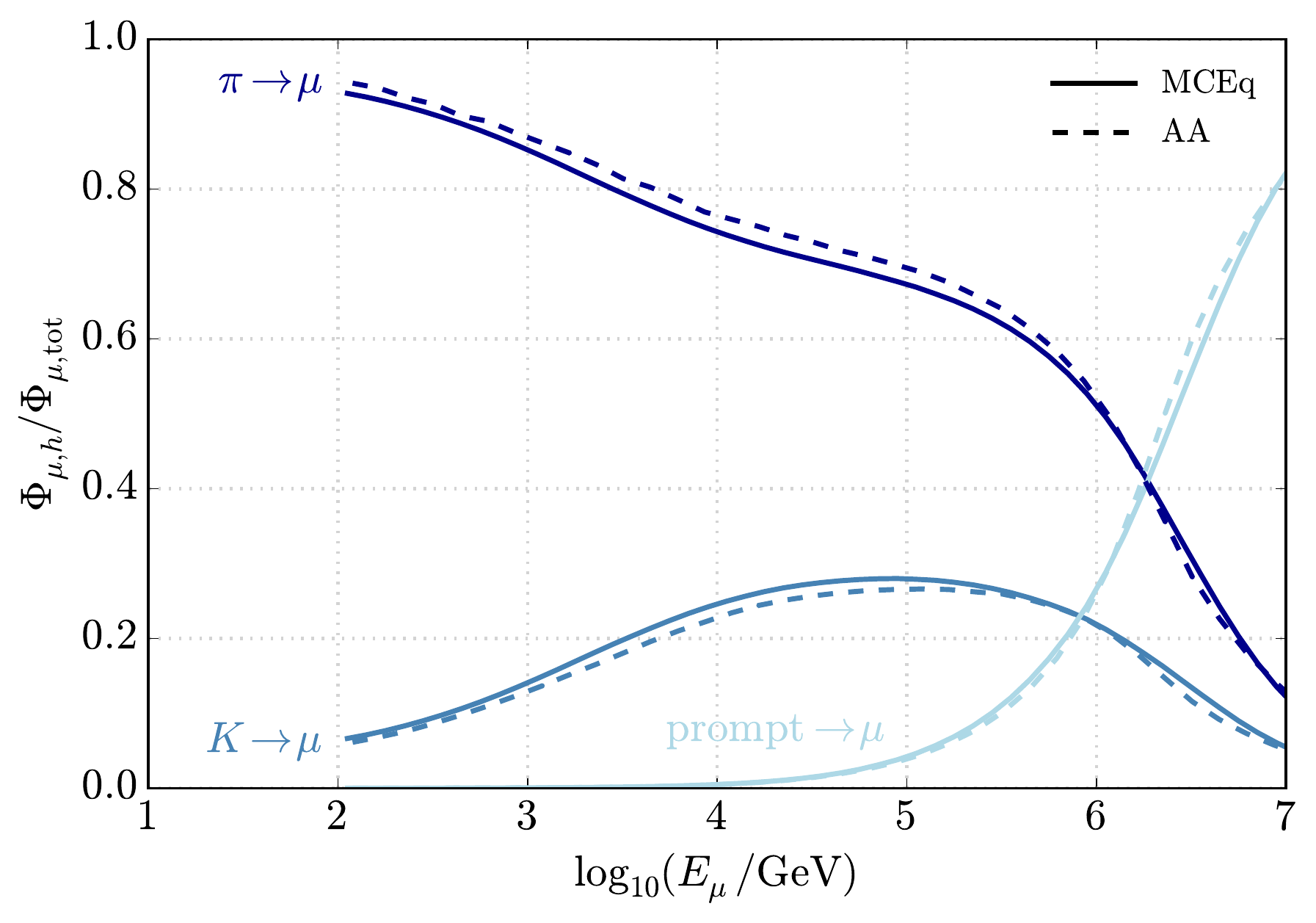}\includegraphics[width=0.5\textwidth]{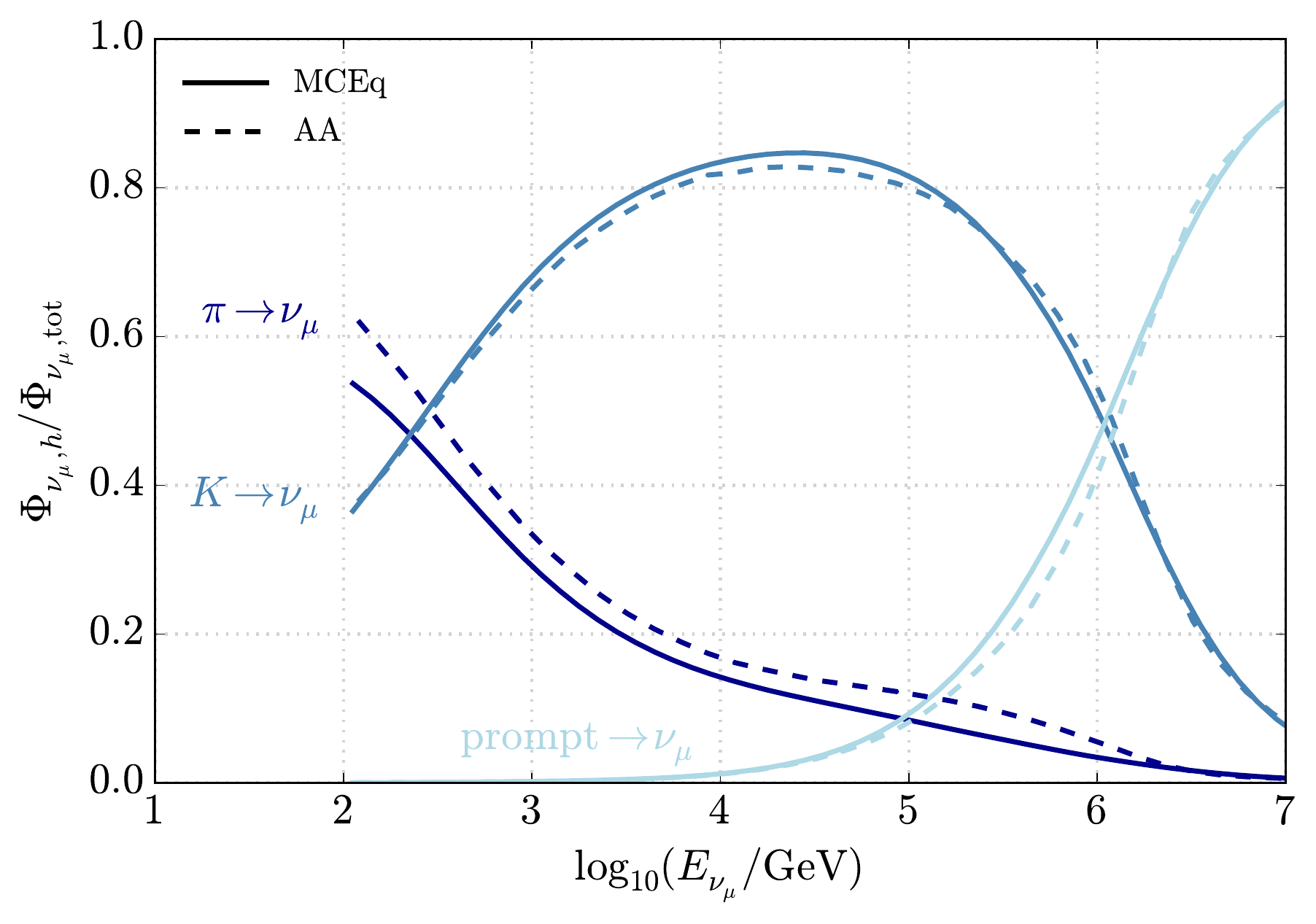}\vspace{-0.3cm}
\caption{Fraction of muons (left) and muon neutrinos (right) from pions, kaons and prompt sources integrated over zenith angle.}
\label{fig:Kpi}
\end{figure}

The decay moments follow from the two-body decay kinematics
of charged pions and kaons.  In particular,
\begin{equation}
Z_{\pi \mu}(\gamma)\,=\,\frac{(1-r_\pi^{\gamma+1})}{(\gamma+1)(1-r_\pi)}\,=\,\int_{r_\pi}^1 x^\gamma\frac{{\rm d}n_\mu}{{\rm d}x}{\rm d}x
\label{eq:pi2mu}
\end{equation}
and
\begin{equation}
Z_{\pi\mu}(\gamma+1)\,=\,\frac{(1-r_\pi^{\gamma+2})}{(\gamma+2)(1-r_\pi)},
\label{pi2mu-2}
\end{equation}
where $x = E_\mu/E_\pi$, $\gamma$ is the integral spectral index of the cosmic-ray spectrum and $r_\pi = (m_\mu/m_\pi)^2\approx 0.573$.
The forms for two-body decay of charged kaons are the same but with $r_K=(m_\mu/m_K)^2\approx 0.046$. 
The corresponding forms for decay to muon neutrinos are
\begin{equation}
    Z_{M\nu}(\gamma) = \frac{(1-r_M)^{\gamma+1}}{(\gamma+1)(1-r_M)}
    \label{eq:M1}
\end{equation}
and
\begin{equation}
Z_{M\nu}(\gamma+1)\,=\,\frac{(1-r_M)^{\gamma+2}}{(\gamma+2)(1-r_M)}.
\label{eq:M2}
\end{equation}
Because $r_\pi$ is large, the muon carries most of the energy in pion decay, 
while in kaon decay the energy is shared almost equally between the muon and the neutrino.  
As a consequence, the kaon channel becomes the dominant source of $\nu_\mu$ above $\sim 100$~GeV 
where Eq.~(\ref{eq:M2}) applies.  These effects are illustrated in Fig.~\ref{fig:Kpi}.

A standard approximation is to combine the low Eq.~(\ref{eq:lowE}) and high Eq.~(\ref{eq:highE})
energy forms into a single equation of the form~\cite{Gaisser:2016uoy}
\begin{equation}
\frac{{\rm d}N_\ell}{{\rm d}E_\ell} = \frac{N_0(E_\ell)}{1-Z_{NN}}\left\{\frac{A_{\pi\ell}}{1+B_{\pi\ell}\cos\theta \,E_\ell/\epsilon_\pi}+\frac{A_{K\ell}}{1+B_{K\ell}\cos\theta \,E_\ell/\epsilon_K}\right\}.
\label{eq:lepton_approx}
\end{equation}
This formula applies to $\mu^+ +\mu^-$ and to $\nu_\mu + \bar{\nu}_\mu$.  For the treatment
of separate charges we refer to Ref.~\cite{Gaisser:2011cc}.
Prompt leptons can be treated by adding a term of the same form as for pions and kaons
to Eq.~(\ref{eq:lepton_approx}) and using the appropriate decay distributions.  
Typically, however, the analytic approximations contain a limited
number of cascade channels.  For example, Eq.~(\ref{eq:lepton_approx}) neglects production of kaons by pions
(and vice versa).  In addition, since resonance production is not included, the input data
fot the spectrum-weighted moments most include pions and kaons from decay of resonances.

Equation~\ref{eq:lepton_approx} can be generalized to include the slow energy-dependence
of cross sections as well as a smooth bending of the primary spectrum by evaluating 
spectrum weighted moments following the prescription of Thunman, Ingelman and Gondolo~\cite{Gondolo:1995fq}:
\begin{equation}
Z_{NM}(E) = \int_E^\infty\frac{N_0(E^\prime)}{N_0(E)}\frac{\sigma(E^\prime)}{\sigma(E)}\frac{{\rm d}n_{NM}(E^\prime,E)}{{\rm d}E}{\rm d}E^\prime\rightarrow\int_0^1\,x^\gamma\frac{{\rm d}n_{NM}(x)}{{\rm d}x}\,{\rm d}x.
\label{eq:TIG}
\end{equation}
For a power-law primary spectrum with integral spectral index $\gamma$ and constant cross sections the
definition collapses to the standard scaling definition in terms of $x=E/E^\prime$.  The energy dependence of
the decay factors is accounted for by using an energy-dependent
spectral index, $\gamma(E)$.  Note that here 
also, the generalized spectrum weighted moments are evaluated at the energy of the lepton. 

\begin{figure}[t]
\centering
\includegraphics[width=1.\textwidth]{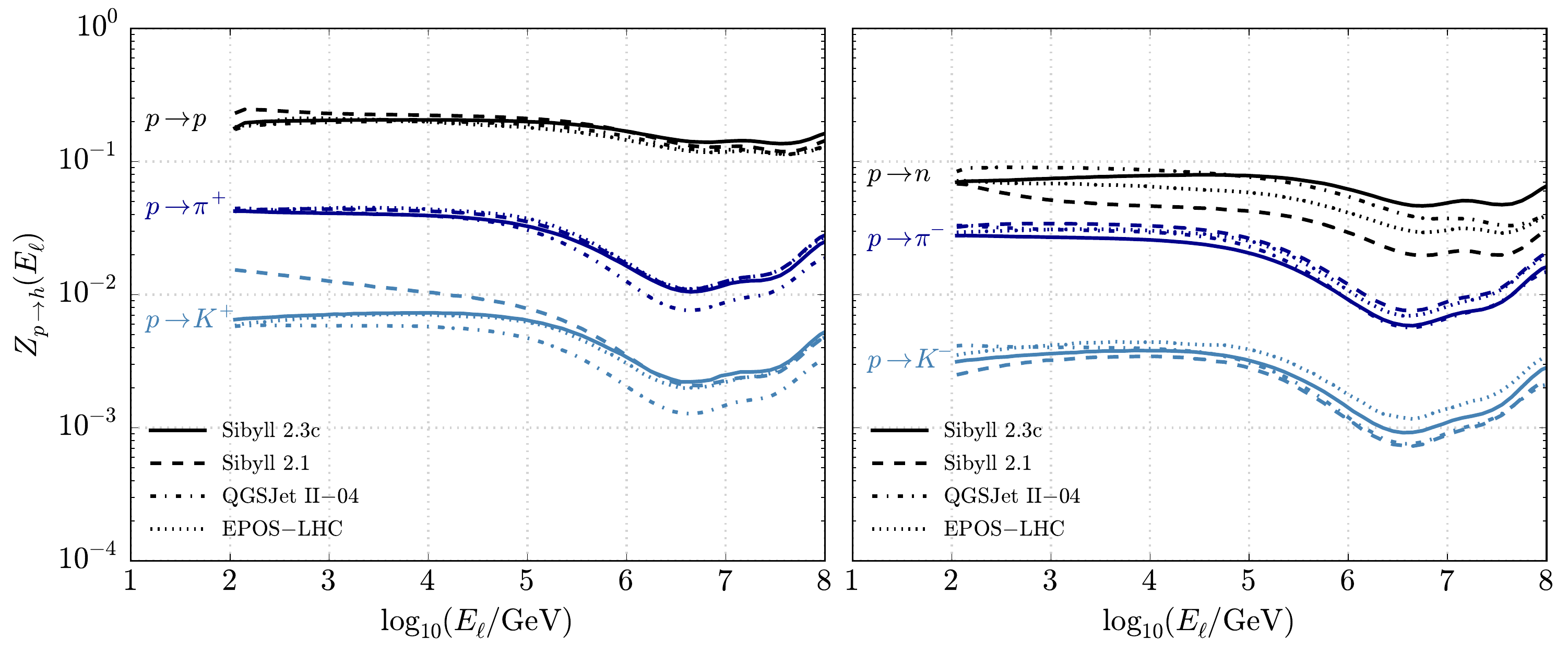}\vspace{-0.3cm}
\caption{Energy-dependent Z-factors calculated with various hadronic models for the H3a nucleon spectrum~\protect\cite{Gaisser:2011cc}.  Left: positive secondaries; Right: neutrons and negative mesons.}
\label{fig:ZofE}
\end{figure}

The next section discusses atmospheric leptons from $100$~GeV to $> 10$~PeV.  For such high-energy
leptons $\gg 100$TeV it is necessary to use the energy-dependent version of the Z-factors
defined in Eq.~(\ref{eq:TIG}) to take account of the knee in the primary spectrum in the PeV range.  
Fig.~\ref{fig:ZofE} compares the Z-factors for the four interaction models used for comparison in
this paper ({\sc Sibyll}-2.3c~\cite{Riehn:2017mfm}, {\sc Sibyll}-2.1~\cite{Ahn:2009wx},
 {\sc EPOS-LHC}~\cite{Pierog:2013ria} and QGSJetII-04~\cite{Ostapchenko:2013pia}).

The Matrix Cascade Equation ({\sc MCEq}) program\footnote{https://github.com/afedynitch/MCEq}~\cite{MCEq,Fedynitch:2018vfe} solves Eq.~(\ref{eq:MasterEqn}) by 
starting with a nucleon of energy $E_0$ and integrating by matrix multiplication at
each step of ${\rm d}X$, using one of the standard hadronic interaction models
to calculate production of particles.  It follows 65 particle types with an energy grid of eight
bins per decade of energy, and it includes energy loss for charged particles.  Thus the
two main limitations of the approximate formulas (limited channels and the
interpolation between low and high energy) are absent.  On the other hand, the
 analytic approximations are parameterised directly in terms of the underlying 
 physical processes.  In this paper we compare several results of the analytic approximations 
 (hereafter AA)
 with those of {\sc MCEq}.  The goal, using {\sc MCEq} as the standard, is to compare errors and limitations
 of the approximate formulas with differences between the different cosmic-ray event generators.
\begin{figure}[b]
\centering
\includegraphics[width=0.543\textwidth]{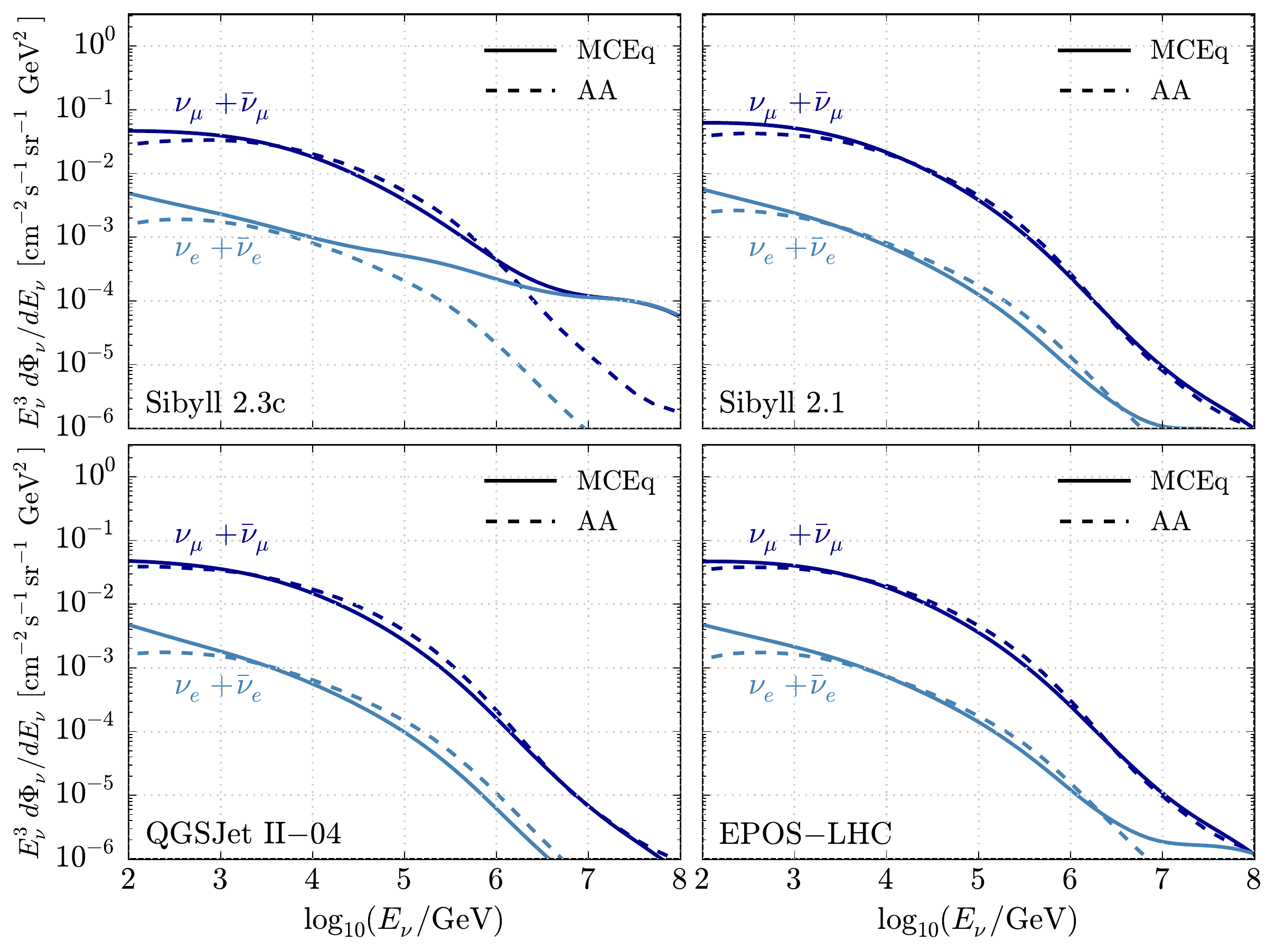}\,\includegraphics[width=0.449\textwidth]{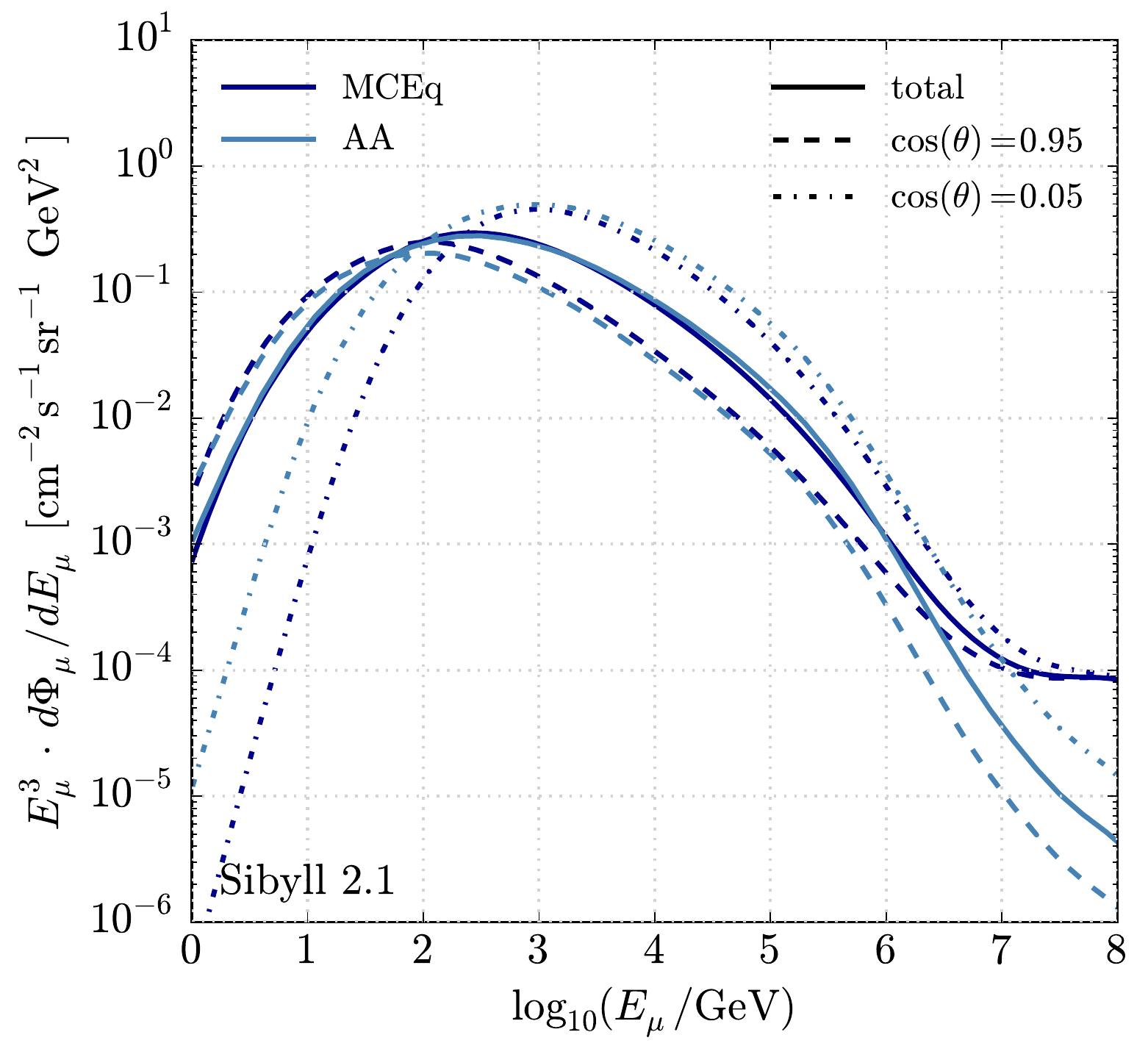}\vspace{-0.2 cm}
\caption{Left: Comparison of neutrino fluxes averaged over zenith angle for four hadronic interaction models (solid, {\sc MCEq}; dashed AA) 
with the H3a spectrum of Ref.~\protect\cite{Gaisser:2011cc}.
Right: The calculation with {\sc Sibyll}-2.1 for muons, also shown for $\cos\theta=0.95$ and $\cos\theta=0.05$ separately.}
\label{fig:4neutrino}
\end{figure}

 \vspace{-.5cm}
 \section{Comparison of lepton fluxes}
 \label{sec:compare}
 In this section we show examples of fluxes of muons and neutrinos.  In Fig.~\ref{fig:4neutrino}
 the comparisons between {\sc MCEq} and AA for neutrinos are shown for {\sc Sibyll}-2.3c~\cite{Riehn:2017mfm}, {\sc Sibyll}-2.1~\cite{Ahn:2009wx},
 {\sc EPOS-LHC}~\cite{Pierog:2013ria} and QGSJetII-04~\cite{Ostapchenko:2013pia}.  The AA calculations do not include prompt neutrinos,
 which are present in {\sc Sibyll}-2.3c, but not in the other models.  In addition, AA does not include
 neutrinos from decay of muons, which are significant below a TeV, especially for $\nu_e$.  The
 right panel of the figure compares {\sc MCEq} and AA averaged over zenith angle for {\sc Sibyll}-2.1, 
 which does not include charm.  However,
 it does include prompt muons from the $\mu^+ + \mu^-$ branch of neutral vector mesons decay,
 which accounts for the rise in the {\sc MCEq} muon flux at high energy.  The AA muon calculation
 accounts for muon energy loss with an approximation due to Lipari~\cite{Lipari:1993hd}
and described in  Ref.~\cite{Gaisser:2016uoy}.  The excess of {\sc MCEq} in electron neutrinos towards low energies is the contribution from decays of secondary muons, a process that is currently missing in the AA.

\begin{figure}[t]
\centering
\includegraphics[width=0.45\textwidth]{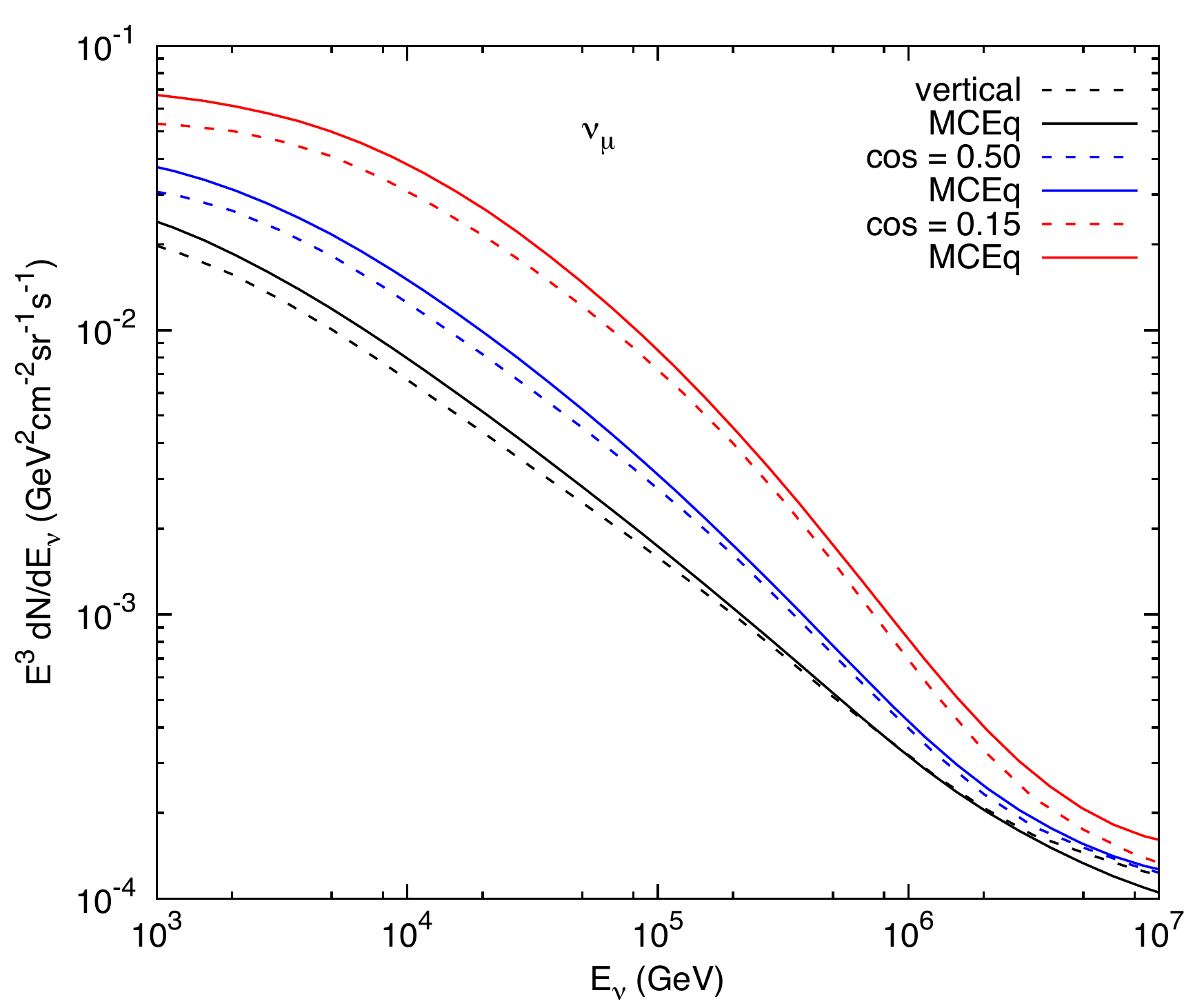}\includegraphics[width=0.45\textwidth]{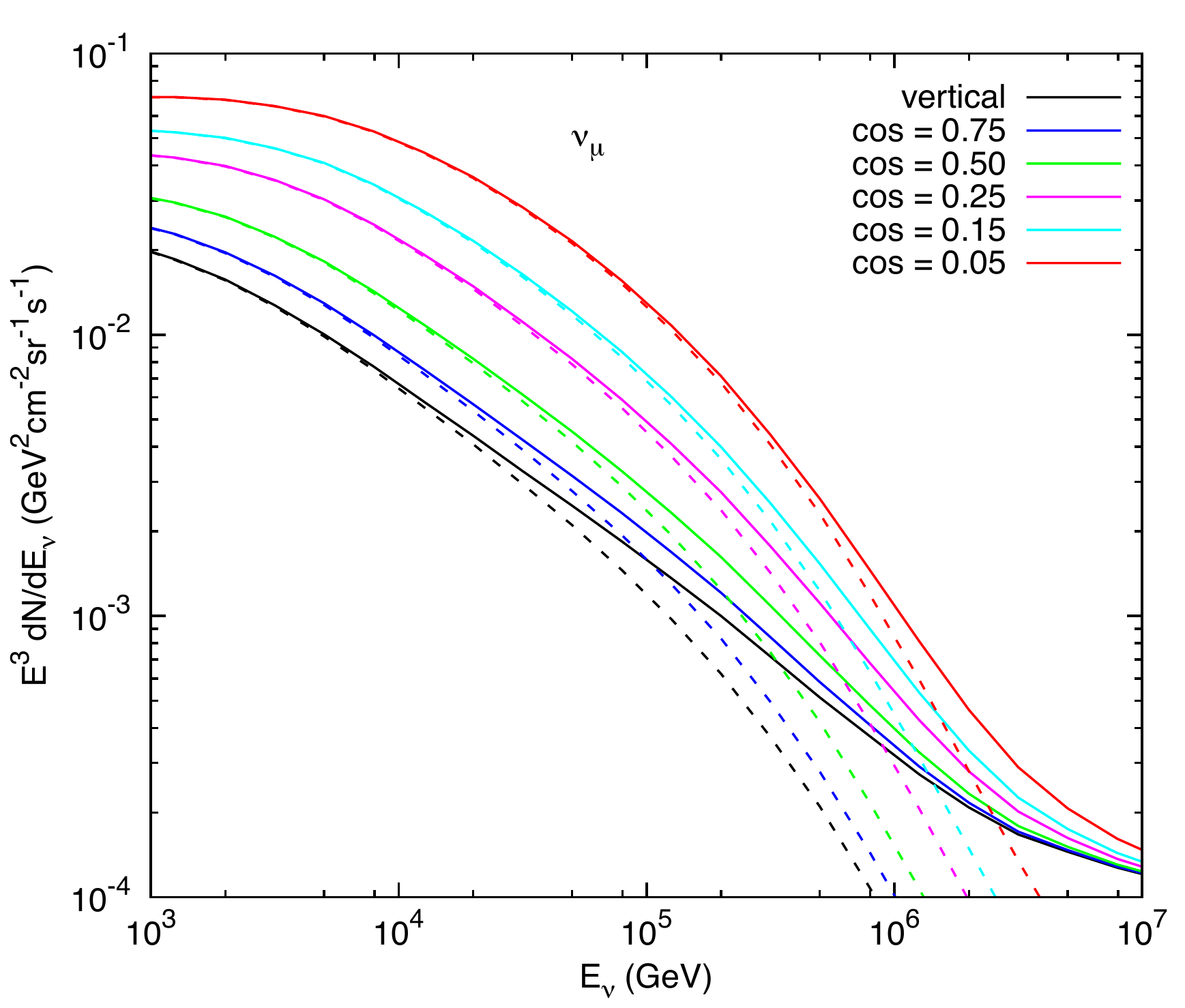}\vspace{-0.3cm}
\caption{Left: Comparison of muon neutrino fluxes $\nu_\mu + \bar{\nu}_\mu$ using {\sc Sibyll}-2.3c and the H3a flux (solid, {\sc MCEq}; dashed AA); Right: Flux of $\nu_\mu + \bar{\nu}_\mu$
for several zenith angles (AA conventional neutrinos with broken lines).}
\label{fig:compare_nu}
\end{figure}

Fig.~\ref{fig:compare_nu} illustrates the fact that the flux of high-energy atmospheric neutrinos
 increases significantly toward the horizon as a consequence of the ''secant$\theta$'' effect
 from the $1/\cos\theta$ factor in Eq.~(\ref{eq:lepton_approx}).  The prompt contribution
 to neutrino event rates is much smaller than the impression given by the amplification
 by $E^3$ in the plot.  The flux of atmospheric neutrinos increases by almost an order of magnitude from vertical to horizontal in the 10 to 100~TeV energy range, in contrast to the isotropic expectation for a diffuse astrophysical flux. 

\begin{figure}[b]
\centering
\includegraphics[width=0.5\textwidth]{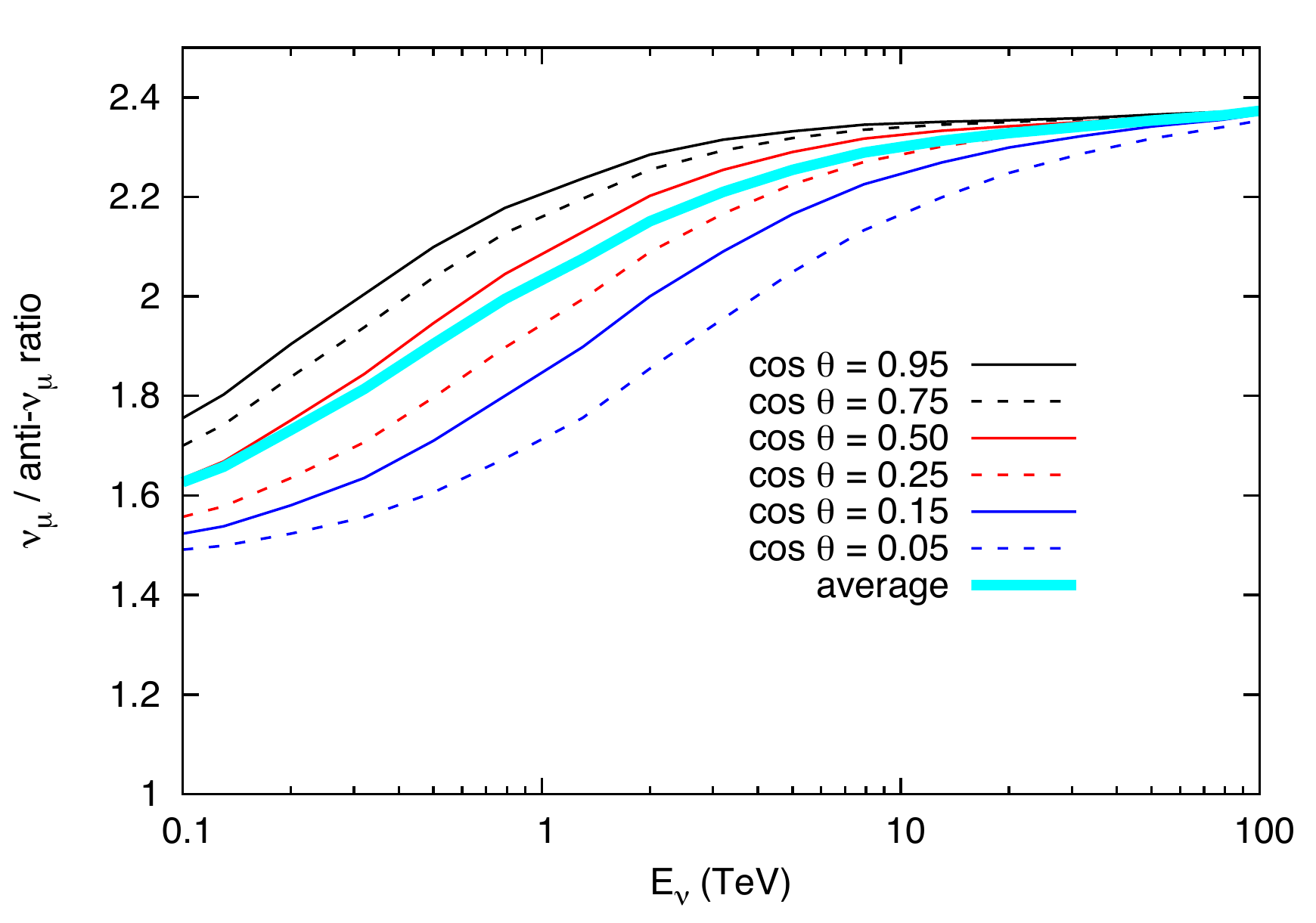}\includegraphics[width=0.5\textwidth]{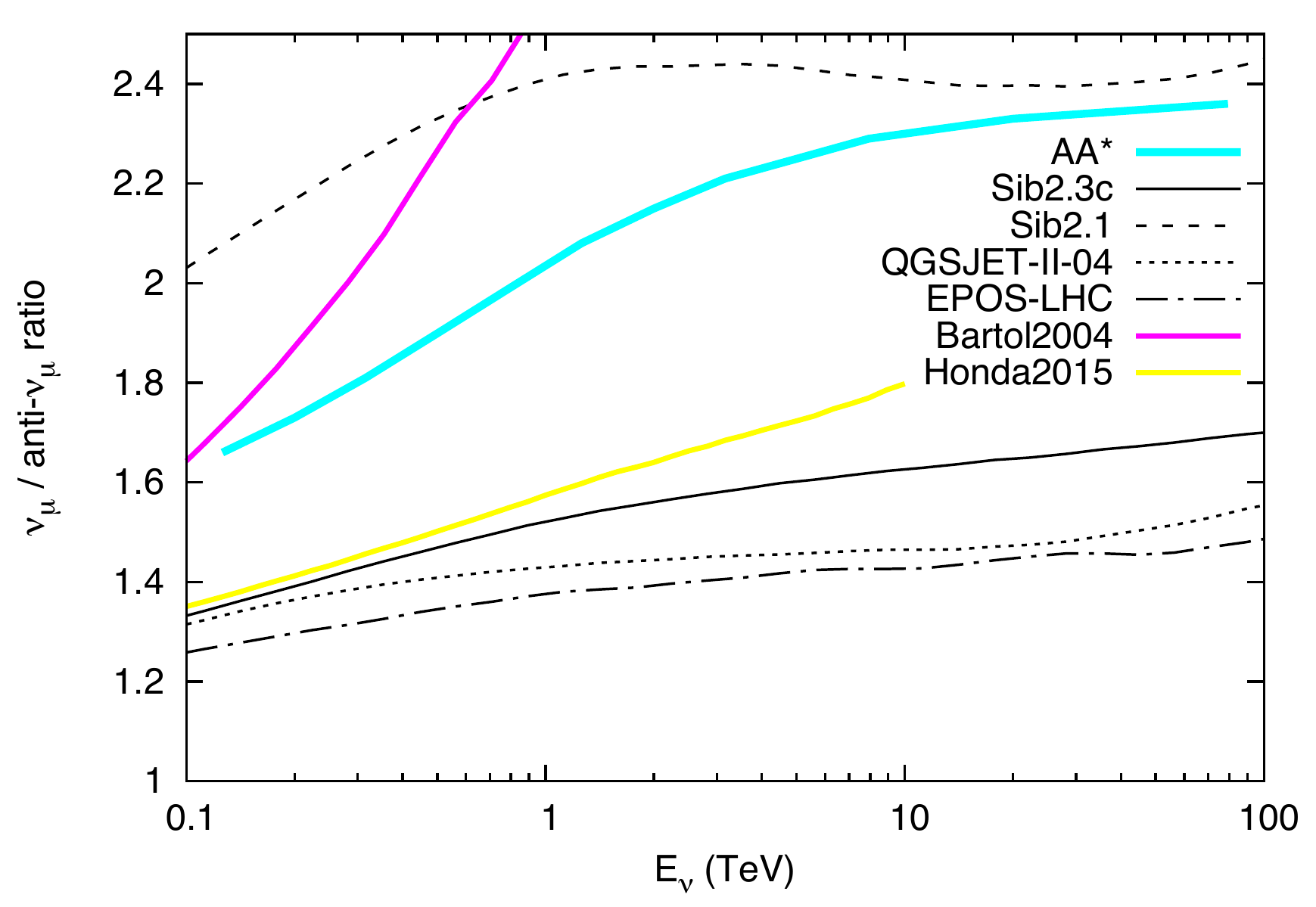}\vspace{-0.3cm}
\caption{Left: The predicted ratio $\nu_\mu/\bar{\nu}_\mu$ for a range of $\cos\theta$ ; Right: Ratios averaged over zenith angle from six specific interaction models compared to the analytical approximation using OPERA parameters (AA*), as described in the text.}
\label{fig:nu_nubar}
\end{figure}

\vspace{-.5cm}
\section{Charge ratio}
\label{sec:charge_ratio}

The increasing importance of the kaon channel at high energy (Fig.~\ref{fig:Kpi}) leads to
an increase in the muon charge ratio at TeV energies.  This is a consequence of the fact
that the charge ratio of kaons is larger than that of pions.  The muon charge ratio also depends
on the parameter $\delta_0 = (p-n)/(p+n)$, the proton excess in the spectrum of primary nucleons.
The OPERA Experiment~\cite{Agafonova:2014mzx} fit their measurement of the muon charge ratio 
using the parameterisation of
Ref.~\cite{Gaisser:2011cc} with the two main parameters, $\delta_0$ and $Z_{pK^+}$, adjusted to
fit their data.  Here we show in the left panel of Fig.~\ref{fig:nu_nubar} the predicted $\nu_\mu/\bar{\nu}_\mu$ ratio
calculated using the OPERA parameters (AA*). The right panel compares {\sc MCEq} calculations of the $\nu_\mu/\bar{\nu}_\mu$ ratio for the four event generators for air showers as well as two Monte Carlo calculations at low energy, Bartol2004~\cite{Barr:2004br} and Honda2015~\cite{Honda:2015fha}.  The left plot illustrates how the ratio evolves with energy as a function of zenith angle, which follows from the $E_\nu\cos\theta$ dependence in Eq.~(\ref{eq:lepton_approx}).

\section{Seasonal variation of muons}
\label{sec:seasonal}

\begin{figure}[t]
\centering
\includegraphics[width=1.0\textwidth]{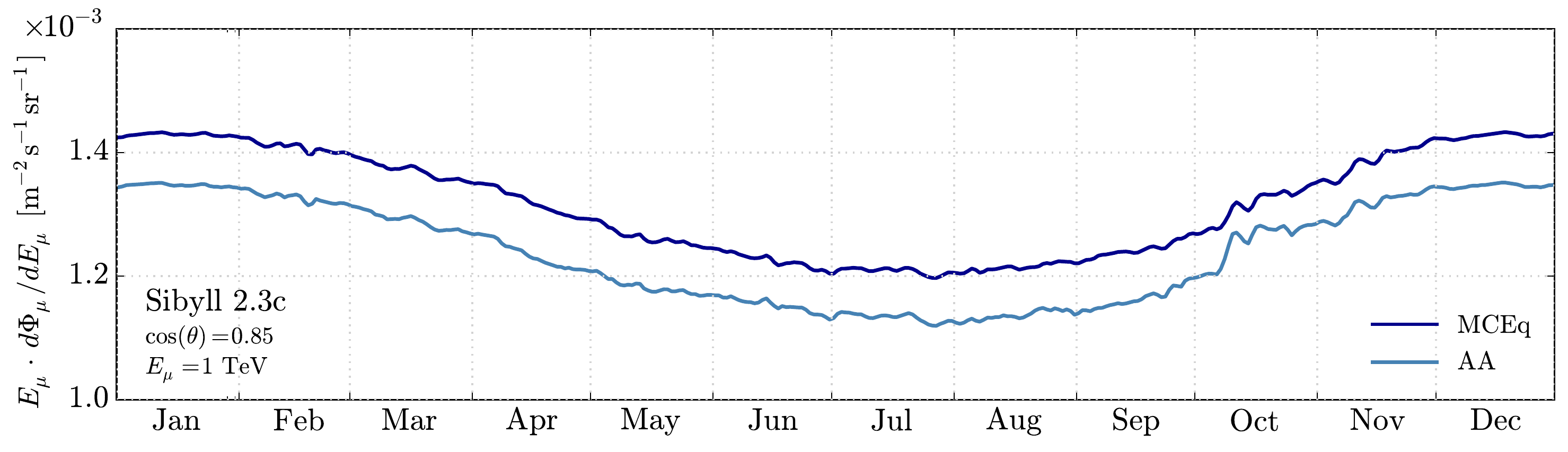}\vspace{-0.3cm}
\caption{Atmospheric muon rate at $E_\mu=1\,\mathrm{TeV}$ and $\cos\theta=0.85$ from {\sc MCEq} and AA calculations for each day of the year 2012 using {\sc Sibyll}-2.3c and the H3a cosmic ray flux.}
\label{fig:seasonal}
\vspace{-0.1 cm}
\end{figure}

A traditional measurement for any underground cosmic-ray detector is the seasonal variation of the muon rate.  The range of depths of various detectors corresponds to an energy range from below ($100$~GeV) to several TeV~\cite{Adamson:2015qua}.  The 
temperature dependence enters through the pion and kaon critical energies, which are proportional to the absolute temperature as a function of atmospheric depth from 
Eq.~(\ref{eq:critical}).  The full analysis for the IceCube detector at the South Pole
is described in Ref.~\cite{Tilav:2019icrc} presented at this conference.   The extreme temperatures at the South Pole make this an interesting test of the comparison between AA and {\sc MCEq}, as illustrated in Fig.~\ref{fig:seasonal}.

The full calculation for the South Pole~\cite{Tilav:2019icrc} requires accounting for the energy-dependence of the acceptance for muons in IceCube as well as the energy dependence of muon production.  Here we show the calculated muon rate for $E_\mu=1$~TeV for each day in 2012, which is independent of IceCube response to muons.  Using the same temperature data for the South Pole from the AIRS Satellite~\cite{AIRS}, both the overall seasonal variation as well as short term features are reproduced well.  The AA normalization is about 5\% lower than that of {\sc MCEq}, and the short-term features are sharper in AA.

\section{Conclusion}
\label{sec:conclusion}
Lepton fluxes calculated with the analytic approximations generally agree within 5-20\% depending on energy and angle (e.g. Figs.~\ref{fig:compare_nu} and~\ref{fig:seasonal}). Differences among the post-LHC models are in some cases larger as can be seen from Fig.~\ref{fig:ZofE}.  The largest difference is between {\sc Sibyll}-2.3c and {\sc QGSJetII-04} for which $K^+$ differ by $\approx 40$\% while $K^-$ are almost the same.  The situation is similar for pions in that $\pi^-$ are comparable in the two models and $\pi^+$ are $\approx 15$\% higher in {\sc Sibyll}-2.3c. {\sc EPOS-LHC} is intermediate.  The charge ratio differences are also reflected in the differences among the models for the $\bar{\nu}_\mu/\nu_\mu$ in the right panel of Fig.~\ref{fig:nu_nubar}.  Interestingly, the prediction for the $\bar{\nu}_\mu/\nu_\mu$ ratio based on parameters of the muon charge ratio measurement from OPERA~\cite{Agafonova:2014mzx} is higher than all the post-LHC models.

\end{document}